\documentclass{aa}  

\usepackage{graphicx}
\usepackage{txfonts}
\usepackage{lscape}
\usepackage[flushleft]{threeparttable}
\usepackage{xcolor}
\begin{document} 

   \title{A new method of measuring Forbush decreases}

\author{M. Dumbovi\'c \inst{1}
\and L. Kramari\'c\inst{1,2}
\and I. Benko \inst{1,3}
\and B. Heber \inst{4}
\and B. Vr\v{s}nak \inst{1}
          }

\institute{
University of Zagreb, Faculty of Geodesy, Hvar Observatory, Zagreb, Croatia\\
\email{mdumbovic@geof.hr}
\and University of Zagreb, Faculty of transport and traffic sciences, Zagreb, Croatia
\and Swedish Institute of Space Physics, Uppsala, Sweden
\and Christian-Albrechts University in Kiel, Department of Extraterrestrial Physics, Kiel, Germany
}


 
  \abstract
   {Forbush decreases (FDs) are short-term depressions in the galactic cosmic ray (GCR) flux and one of the common signatures of coronal mass ejections (CMEs) in the heliosphere. They often show a two-step profile, the second one associated with the CME's magnetic structure. This second step can be described by the recently developed analytical FD model for flux ropes (FRs) --ForbMod.}
   {The aim of this study is to utilise ForbMod to develop a best-fit procedure to be applied on FR-related FDs as a convenient measurement tool. Our motivation is to develop a best-fit procedure that can be applied to a data series from an arbitrary detector. Thus, {the basic procedure would facilitate measurement estimation of the magnitude of the FR-related FD}, with the possibility of being adapted for the energy response of a specific detector for a more advanced analysis.}
   {
The non-linear fitting was performed by calculating all possible ForbMod curves constrained within the FR borders to the designated dataset and minimising the mean square error (MSE).   
 In order to evaluate the performance of the ForbMod best-fit procedure, we used synthetic measurements produced by calculating the theoretical ForbMod curve for a specific example CME and then applying various effects to the data to mimic the imperfection of the real measurements. We also tested the ForbMod best-fit function on the real data, measured by detector F of the SOHO/EPHIN instrument on a sample containing 30 events, all of which have a distinct FD corresponding to the magnetic obstacle (MO). The extraction of FD profiles (from the onset to the end) was performed manually by an observer, whereby we applied two different versions of border selection and assigned a quality index to each event.}
   {We do not find notable differences between events marked by a different quality index. For events with a selection of two different borders, we find that the best fit applied on extended interplanetary coronal mass ejection (ICME) structure borders results in a slightly larger MSE and differences compared to the traditional method due to a larger scatter of the data points. We find that the best-fit results can visually be categorised into six different FD profile types. Although some profiles do not show a visually pleasing FD, the ForbMod best-fit function still manages to find a solution with a relatively small MSE. }
   {Overall, we find that the ForbMod best-fit procedure performs similar to the traditional algorithm-based observational method, but with slightly smaller values for the FD amplitude, as it's taking into account the noise in the data. Furthermore, we find that the best-fit procedure has an advantage compared to the traditional method as it can estimate the FD amplitude even when there is a data gap at the onset of the FD.}


\maketitle
%
\section{Introduction}
\label{intro}

Forbush decreases (FDs) are short-term depressions in the galactic cosmic ray (GCR) flux that result from the interaction of GCRs with coronal mass ejections (CMEs). They are thus one of the common signatures of CMEs in the heliosphere \citep[see overviews by e.g.][]{lockwood71,cane00,belov09}. Detecting CMEs at various different locations in the heliosphere is crucial for the analysis of their evolution, as well as the validation of the CME propagation and evolution models. They are typically observed by in situ detectors that measure magnetic field and plasma properties \citep[e.g.][]{zurbuchen06,kilpua17}; however, these are not available as frequently as particle detectors that can detect FDs. Therefore, a better understanding of FDs can lead to CME analysis where other types of observation are not possible.

\begin{figure*}
\centerline{\includegraphics[width=0.8\textwidth]{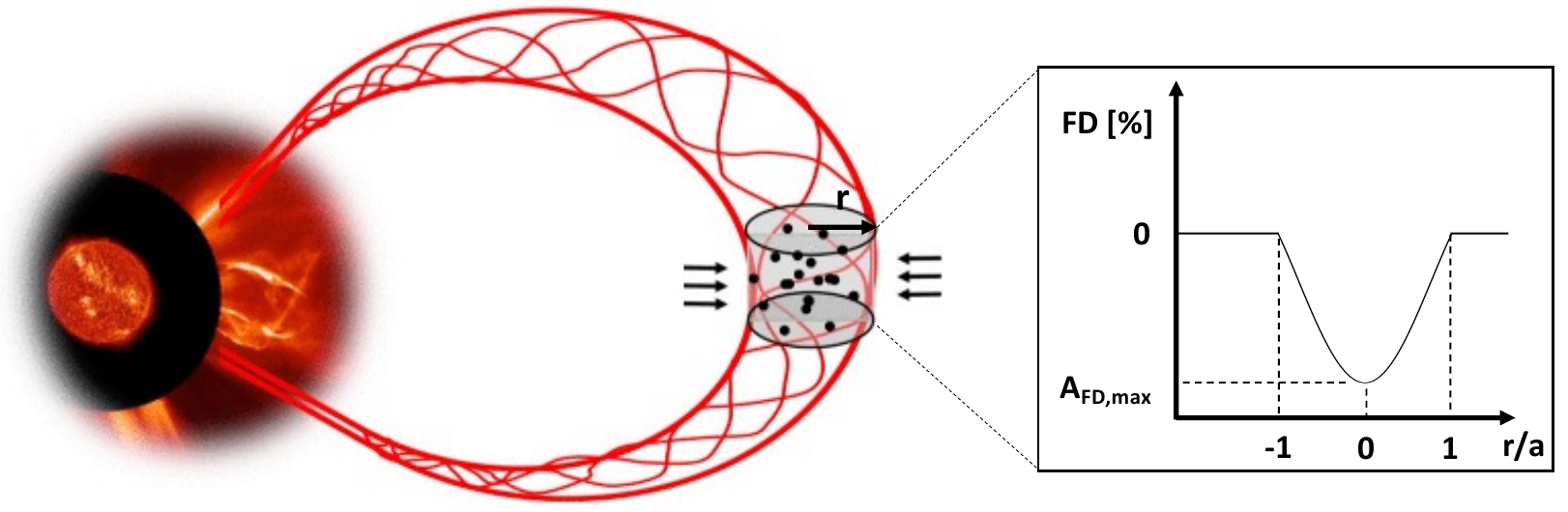}}
\caption{Sketch of the ForbMod model for FR-related FDs. The GCRs diffuse into the self-similarly expanding FR (r/a=const.), which is initially empty at its centre (r/a=0), where a(t) is the FR radius and r(t) the radial distance from the centre of the FR to its borders. As a result, at an arbitrary time after the FR eruption from the Sun (i.e. at an arbitrary heliocentric distance), a distribution of GCRs inside the FR is observed, given by the Bessel function. The relative GCR density (FD[\%]) at the borders of the FR is 0\%, whereas in the centre of the FR it is given by the amplitude of the depression ($A_{\mathrm{FD,max}}$). For model details, see the main text.}
\label{fig_forbmod}
\end{figure*}

FDs caused by interplanetary coronal mass ejections (ICMEs) often show a two-step profile, one associated with the turbulent and compressed shock-sheath region ahead of the CME magnetic structure and the other with the CME magnetic structure itself \citep{cane00}. Because of their different physical properties, the mechanism through which these two regions produce the depression is different, and thus should be modelled differently. The current paradigm is that the CME magnetic structure is a flux rope (FR), a structure with field lines helicoidally winding around a central axis \citep[e.g.][]{zhang21}. It was first proposed by \citet{morrison56} that FDs might be caused by clouds of closed magnetised plasma, whereby they are initially empty of particles and slowly fill as they propagate through the interplanetary space. This approach has since been utilised to explain FR-associated FDs, where the particles are allowed to enter via ``cross-field'' transport; in other words, perpendicular diffusion \citep[e.g.][]{cane95,quenby08}. In addition to the particle diffusion, another important contribution comes from the expansion of the FR. Namely, as they propagate through the IP space, CMEs expand due to the pressure imbalance \citep[e.g.][]{klein82,demoulin09}, where, consequently, as the size of the magnetic structure increases, its magnetic field weakens \citep{bothmer98,demoulin08,vrsnak19}. It was first proposed by \citet{laster62} and \citet{singer62} that expansion is needed to explain the observational properties of FDs. More recently, the diffusion-expansion approach was applied by \citet{munakata06} and \citet{kuwabara09} in a numerical model and by \citet{subramanian09} and \citet{arunbabu13} in an analytical model, considering isotropic self-similar expansion. In the recent FD analytical model ForbMod \citep{dumbovic18b}, different types of expansion are considered for the expansion-diffusion approach, constrained by the observational ranges of the expansion factors \citep{gulisano12}. ForbMod was advanced to take into account the energy dependence of the detector when comparing the model results to measurements \citep{dumbovic20}, since FDs are energy-dependent phenomena \citep[i.e. their amplitudes depend on the energy range and cut-off of the detector which detects them][]{cane00}. This facilitates direct comparison of modelled and observed FDs, as was shown by, for example, \citet{forstner21}.

It should be noted that there are other approaches to FD modelling in the FR, such as perpendicular transport by guiding centre drifts \citep[e.g.][]{krittinatham09,tortermpun18}, full trajectory integration using CME FR-type models \citep{petukhova19b}, or CME magnetic field reconstructions from in situ measurements \citep{benella20}, as well as describing FDs via the change in the single GCR spectrum modulation parameter attributed with a CME \citep{guo20}. However, a recent study by \citet{laitinen21}, in which they performed full-orbit particle simulations, with the interface between the external interplanetary magnetic field and the isolated FR field lines modelled analytically, revealed that the transport around the FR border is fast compared to diffusive radial propagation within the FR. As a result, the propagation of GCRs into the FR can be modelled as diffusion into a cylinder. Moreover, the diffusive approach is further supported by most recent observations \citep[see e.g. recent studies by ][]{janvier21, davies23}.

The aim of this study is to utilise ForbMod to develop a best-fit procedure that can be applied to FR-related FDs as a convenient measurement tool. Our motivation is to develop a best-fit procedure that can be applied to a data series from an arbitrary detector. Thus, the basic procedure would facilitate measurement estimation of the onset, duration, and magnitude of the FR-related FD, with the possibility of being adapted for the energy response of a specific detector for a more advanced analysis. {It should be noted that in many cases the onset of the FR-related FD does not necessarily correspond to the onset of the FD itself, as the latter might correspond to the preceding shock-sheath region. In such cases we refer to the onset of the second decrease as the onset of FD (or FR-related FD).} In section \ref{data} we describe the mathematical formalism of the best-fit procedure as well as two samples used in the study to test its performance. We first used synthetic measurements to test the performance of the best-fit procedure under controlled conditions. Next we used a real sample of events detected by the F-detector of the Electron Proton Helium INstrument instrument aboard the SOHO spacecraft \citep[SOHO/EPHIN][]{muller-mellin95}. The results are presented and discussed in section \ref{results}.

\section{Methods and data}
\label{data}
\subsection{ForbMod best-fit procedure}
\label{best-fit}

ForbMod is an analytical diffusion-expansion model that assumes that the CME magnetic structure (i.e. FR) can locally be represented by a straight cylinder, which is initially empty at its centre. The FR is assumed to be propagating at a constant speed, expanding self-similarly, and the particles can only enter it via perpendicular diffusion. It is assumed that GCRs are mostly protons. The time-dependent transport equation with these assumptions simplifies to a time-dependent radial diffusion of protons into an infinite straight cylinder. The corresponding solution is a combination of the radial-dependent solution, which is given by the cylindrical Bessel function of the zero-order, and the time-dependent solution, which is given by the exponential function \citep[for details on the model see][]{dumbovic18b}. The amplitude of the FD, defined as normalised to border conditions, depends on the diffusion rate of particles into the FR. The diffusion rate is governed by the behaviour of the magnetic field, which is generally decreasing due to expansion. Thus, the FD amplitude is expected to decrease with diffusion time (i.e. helisopheric distance). The sketch visualising ForbMod is given in Figure \ref{fig_forbmod}. The diffusion rate of particles also depends on the particle energy, as higher-energy particles enter the FR more easily. Therefore, it is expected that the FD amplitude is energy-dependent and that it should appear differently in detectors with a different energy response \citep[for details see][]{dumbovic18b,dumbovic20}. 

\begin{figure*}
\centerline{\includegraphics[width=0.9\textwidth]{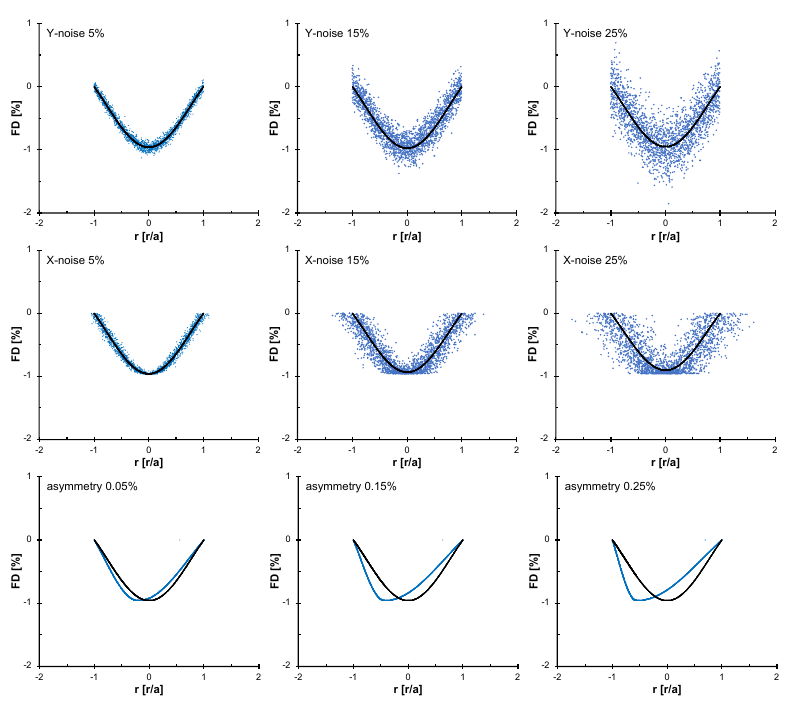}}
\caption{Effects applied to the data to mimic the “imperfection” of the real measurements: Y-noise (upper panels), X-noise (middle panels), and asymmetry (lower panels) for three different levels of noise and asymmetry, respectively. The original theoretical curve is given in black, whereas the data with the applied effect are coloured blue. For details, see the main text.}
\label{fig_synth1}
\end{figure*}

\begin{figure*}
\centerline{\includegraphics[width=0.9\textwidth]{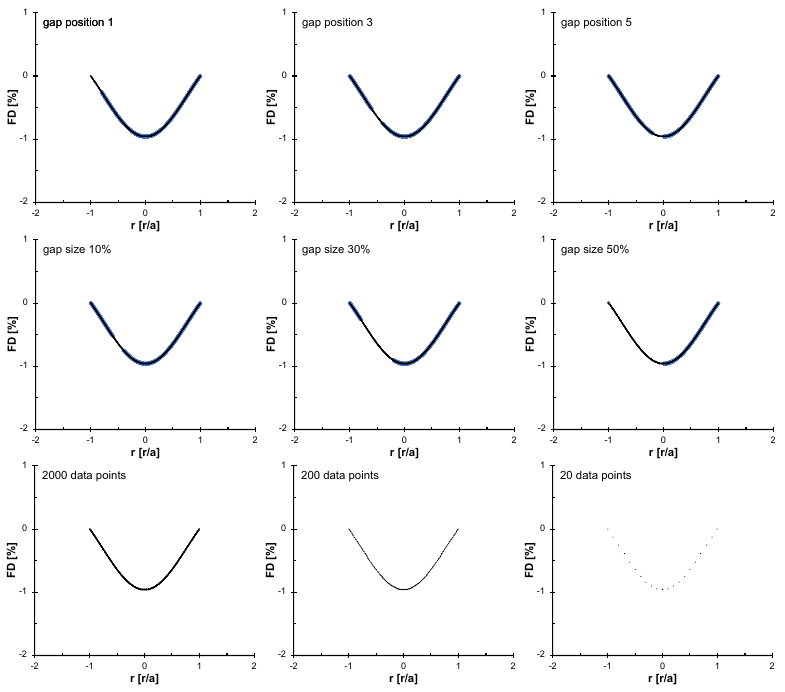}}
\caption{Effects applied to the data to mimic the “imperfection” of the real measurements: data gaps at certain positions (upper panels), data gaps of a certain size (middle panels), and different data resolutions (lower panels). The original theoretical curve is given in black, whereas the data with the applied effect are coloured blue. For details, see the main text.}
\label{fig_synth2}
\end{figure*}

\begin{figure*}
\centerline{\includegraphics[width=0.65\textwidth]{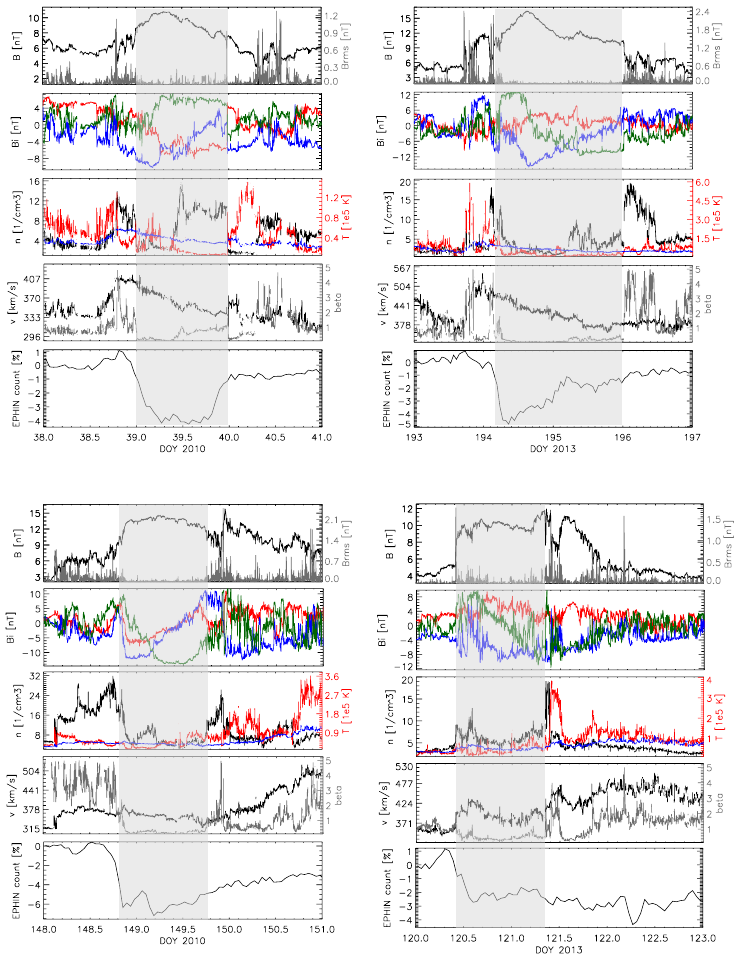}}
\caption{In situ measurements of four different events assigned with four different quality indices, QI4--QI1, from top left to bottom right. Top-to-bottom panels for each event show: 1) magnetic field strength (black, in nT) and its fluctuations (grey, in nT); 2) magnetic field GSE components, x, y, and z, coloured red, blue, and green, respectively (in nT); 3) plasma density (black, in $\mathrm{cm}^{-3}$), temperature (red, in 1e5 K), and expected temperature (blue, in 1e5 K); 4) plasma flow speed (black, in km/s) and beta (grey, non-dimensional); 5) the SOHO/EPHIN F-detector particle count (in \%). The shaded region represents the magnetic obstacle, where the borders were selected according to the low-beta condition.}
\label{fig_qi}
\end{figure*}

\begin{figure}
\centerline{\includegraphics[width=0.45\textwidth]{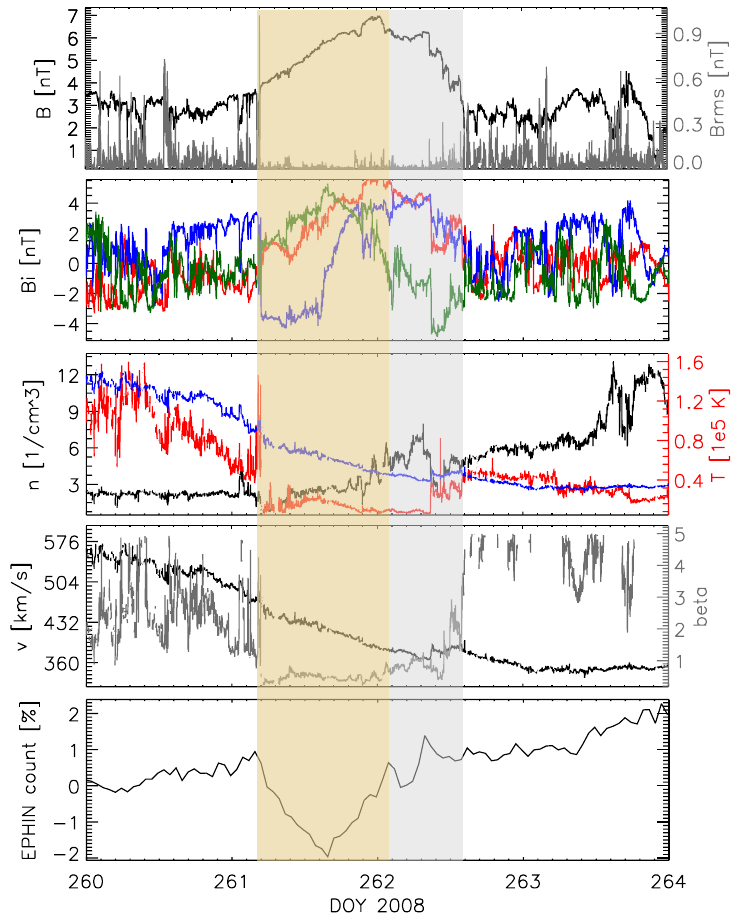}}
\caption{Two different versions of border selection: inner structure (yellow shading) and extended ICME signatures (grey shading). For details, see the main text.}
\label{fig_borders}
\end{figure}

\begin{figure*}
\centerline{\includegraphics[width=0.99\textwidth]{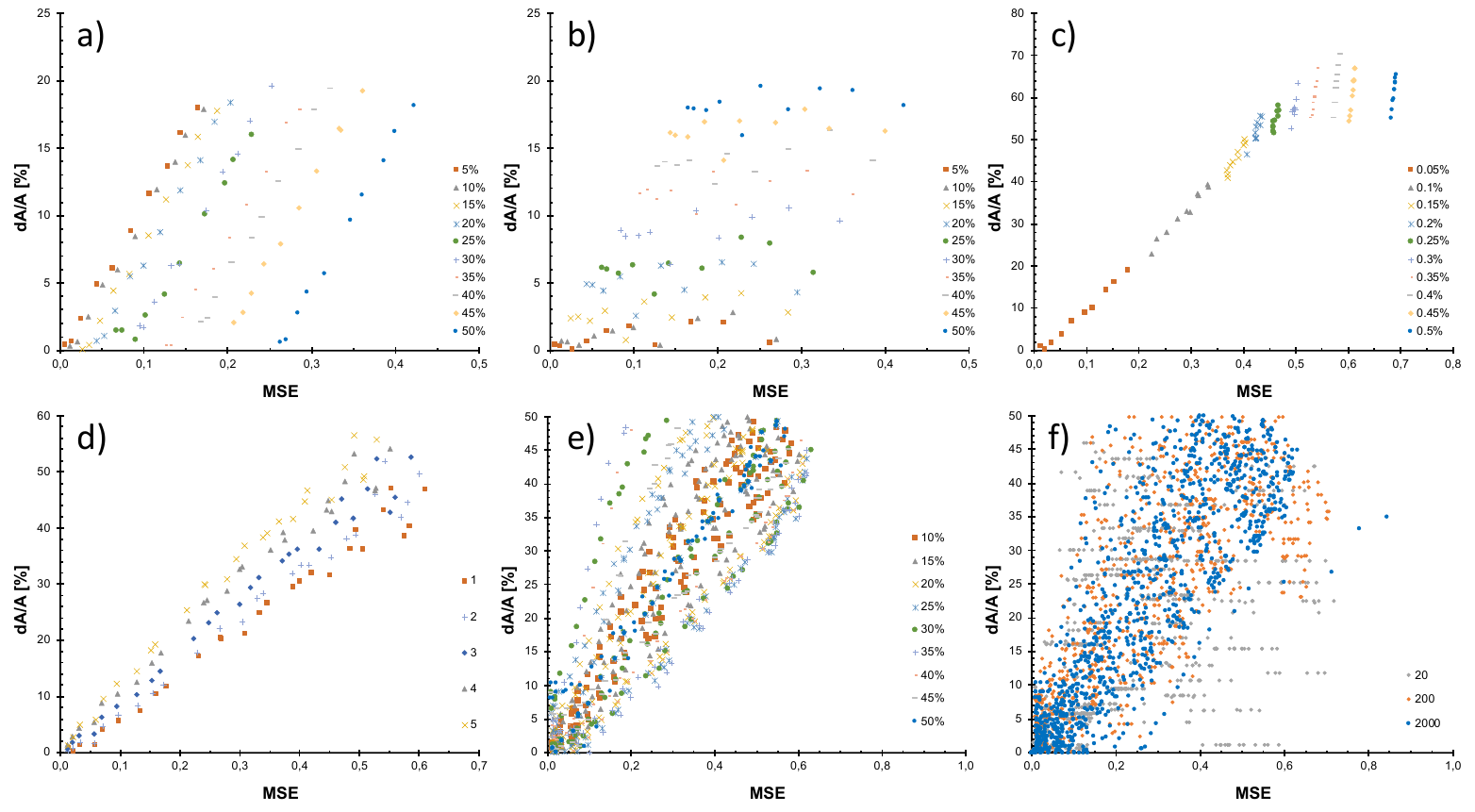}}
\caption{Relative difference of the best-fit FD amplitude to the FD amplitude of the theoretical FD vs goodness of fit for different levels of data imperfection effects: a) Y-noise, b) X-noise, c) asymmetry, d) data gap position, e) data gap size, and f) resolution (for details, see the main text).}
\label{fig_mse}
\end{figure*}

{We developed the ForbMod best-fit procedure by applying a non-linear fitting to the data. For that purpose we first used synthetic measurements, as is described in Section \ref{synthetic}. In general, the non-linear fitting was performed by calculating all possible ForbMod curves constrained within FR borders to the designated dataset and minimising the mean square error, $MSE$}:

\begin{equation}
MSE(y,\hat{y})=\frac{1}{n_{\mathrm{SAMPLES}}}\sum_{i=1}^{n_{\mathrm{SAMPLES}}-1} (y_i-\hat{y}_i)^2\,,
\label{eq_methods-1}
\end{equation}

\noindent where $\hat{y}_i$ is the predicted value of the $i$-th sample, $y_i$ is the corresponding true value, and there are $n_{\mathrm{SAMPLES}}$ samples. The ForbMod best-fit procedure actually performs non-linear fitting of the Bessel function of the zero order (i.e. the radial part of the ForbMod solution) to the designated time series. The input for the procedure is the designated dataset with defined FR borders (i.e. the dataset translated into the FR coordinate system radial direction, with the centre of the FR at $\hat{r}=0$ and FR borders at $\hat{r}=-1$ and $\hat{r}=1$, respectively; $\hat{r}=r/a$, where $a$ is the radius of the FR). The output of the procedure is the best-fitted amplitude of the measured FD, $A_{\mathrm{FR,max}}$, and the corresponding goodness of fit, that is, $MSE$ (as defined by equation \ref{eq_methods-1}). The FD profiles needed to be translated from a timescale into the spatial scale of the FR in order for us to apply the ForbMod best-fit. For that purpose, {we assumed that the spacecraft crosses the full diameter of the FR, which has a circular cross-section. Furthermore, we assumed that the spacecraft trajectory is oriented vertically with respect to the FR axis. We note that with this assumption the model might not realistically capture FRs with orientations deviating significantly from low or high inclinations. Nevertheless, with this assumption,} due to self-similar expansion, the spacecraft observes deceleration of plasma flow across the FR. Thus, the diameter can be given by the equation for accelerated linear motion, $s=0.5at^2 + v_0t$. The borders of the FR were set at [-1,1], per the definition of the ForbMod best-fit function.

\subsection{Synthetic Forbush decrease measurements}
\label{synthetic}

In order to evaluate the performance of the ForbMod best-fit procedure, we used synthetic measurements. Synthetic measurements were produced by calculating the theoretical ForbMod curve for a specific example CME and then applying various effects to the data to mimic the “imperfection” of the real measurements. 

The arbitrary CME example used to produce synthetic measurements \citep[loosely based on event analysed by][]{forstner21} had the following inputs:
\begin{itemize}
\item detector response function for perfect detector: $Y(E)=1$
\item detector cutoff: $E_{\mathrm{cutoff}}=0.05$GeV
\item event month and year: April 2020
\item ICME transit time in hours (i.e. diffusion time), $TT=123.38$h
\item starting distance from the Sun, $R_0=9.54\mathrm{R_{SUN}}$
\item final distance from the Sun, $R_{fin}=1.005$au
\item initial FR radius, $a_0=1.81\mathrm{R_{SUN}}$
\item expansion factor for FR radius, $n_a=1$
\item in situ magnetic field strength, $B=16.2$nT
\item expansion factor for mag. field, $n_B=2$
\end{itemize}

The effects applied to the data to mimic the “imperfection” of the real measurements were:
\begin{itemize}
    \item \textbf{Y-noise:} - random noise in the y direction
    \item \textbf{X-noise:} - random noise in the x direction
    \item \textbf{asymmetry:} - systematic left-shift of the FD minimum in the x direction (to simulate the asymmetric FD profile)
    \item \textbf{data gaps:} - simulating a data gap by removing certain data points around the selected position within the FD
    \item \textbf{low data resolution:} - reducing the number of data points in the FD profile
\end{itemize}

The Y-noise value was calculated and added to each data point of the ForbMod solution in the following way:

\begin{equation}
{y_{i,\mathrm{SYNTH}}=y_{i,\mathrm{ForbMod}}+\mathrm{d}y_i\,\,\,;\,\,\,\mathrm{d}y_i=A_{\mathrm{norm}}\cdot \delta_n\,,}
\label{eq_methods-2}
\end{equation}

\noindent where $y_{i,\mathrm{SYNTH}}$ is the synthetic measurement, $y_{i,\mathrm{ForbMod}}$ is the ForbMod solution data point, $\delta y_i$ is the noise, and $A_{\mathrm{norm}}$ is the randomly selected value from the normal distribution defined by the ForbMod solution amplitude as the mean. $A_{\mathrm{norm}}$ was determined using the Python numpy package function random.normal. $\delta_n$ was an arbitrary percentage applied to the randomly selected $A_{\mathrm{norm}}$. The parameter $\delta_n$ could be varied in order to obtain different percentages (i.e. different levels) of noise. The X-noise value was calculated and applied in the same manner as the Y-noise, only in the x direction. The Y-noise and X-noise of different percentages are shown in the upper and middle panels of Figure \ref{fig_synth1}, respectively.

The asymmetry was introduced by incrementally shifting positive $r$ values corresponding to data points of the ForbMod solution:

\begin{equation}
\delta r_{i,\mathrm{pos}}=r_i+i\cdot\delta_a\,,
\label{eq_methods-3}
\end{equation}

\noindent where $r_i$ is the positive $r$ value corresponding to the $i$-th data point of the ForbMod solution (counted for positive $r$ values only) and $\delta_a$ is a shift percentage between two iterations, which could be varied in order to analyse different percentages (i.e. levels) of asymmetry. After applying the asymmetry, the dataset was re-scaled to be defined on a [-1,1] interval using the sklearn package function minmax\_scale. The asymmetry of different percentages is shown in the lower panels of Figure \ref{fig_synth1}.

The data gaps were introduced by removing a certain number of data points from a certain location within the FD profile. Thus, we could vary both the location of the data gap, as well as its size. This is shown in the example in Figure \ref{fig_synth2}. In the upper panels, a gap 10\% in size (i.e. 200 data points out of 2000) is consecutively shifted from the start of the FD to its minimum (five different positions). In the middle panels of Figure \ref{fig_synth2} various gap sizes are shown, where the gap is centred around the middle of the FD main phase (i.e. the decrease phase). Finally, as the last effect applied to the data to mimic the “imperfection” of the real measurements, we considered data resolution, as is shown in the bottom panels of Figure \ref{fig_synth2}.

\subsection{Event sample and measurements}
\label{sample}

A sample of events was created by associating SOHO/EPHIN detected FDs with in situ detected ICMEs and remotely observed CMEs. We used a list of SOHO/EPHIN detected FDs compiled by \citep{belov21} and a catalogue of near-Earth ICMEs compiled by \citep{richardson10} and regularly updated at CALTECH\footnote{\url{http://www.srl.caltech.edu/ACE/ASC/DATA/level3/icmetable2.htm}}. Since we aimed to use the sample for the follow-up analysis, where a 3D reconstruction using stereoscopic data to obtain CME remote properties would be performed, we constrained our analysis to the time period when at least one STEREO spacecraft was operational, where a reliable CME-ICME association could be made, and where CME signatures in white light coronagraphs were significant enough to perform a 3D reconstruction. The sample contains 30 events in the time period 2007--2019. In this study, we only used the part of the sample containing the interplanetary in situ signatures of ICMEs and the corresponding FD, and thus we only provided a description of the ICME-FD sample in detail, whereas the description of the CME-part of the sample will be given {in a future study}.

Interplanetary coronal mass ejections are roughly simultaneous with FD events \citep{cane96,dumbovic11}, and therefore this association was trivial and performed based on the timing provided by \citet{belov21} or \citet{richardson10}'s catalogues. To analyse the in situ properties, we used one-minute plasma and magnetic field data in the geocentric solar ecliptic (GSE) system provided by the OMNIWeb\footnote{\url{https://omniweb.gsfc.nasa.gov/html/ow_data.html#norm_pla}} database \citep{king05}. We identified ICME signatures using criteria described by \citep[e.g.][]{zurbuchen06,kilpua17}, considering the notification of \citep{rouillard11}, according to which an ICME consists of the compressed, heated, and turbulent shock-sheath region, followed by a magnetic cloud or ejecta. The magnetic cloud is characterised by a low-temperature, low-plasma beta parameter, an enhanced smoothly rotating magnetic field, and an expanding flow speed profile, whereas the term magnetic ejecta is used when these criteria are only partially met. We note that both magnetic cloud and ejecta are nowadays broadly considered as magnetic obstacles \citep[MO,][]{jian06a,nieves-chinchilla18}, and therefore we use this more general term.

Example events are given in Figure \ref{fig_qi}. For each event we analysed: the magnetic field strength, $B$, and its fluctuations, $dBrms$\footnote{the root mean square variation in the vector of the interplanetary magnetic field}; the magnetic field components, $B_x$, $B_y$, and $B_z$; the plasma density, temperature, and expected temperature\footnote{empirical formula based on the speed of the solar wind calculated according to \citet{lopez87} and \citet{richardson95}}; the plasma flow speed and plasma beta parameter; and the SOHO/EPHIN F-detector particle count, $CR\,\,count$\footnote{normalised to the pre-decrease value for each event}. We used hourly averaged detector F data of the SOHO/EPHIN, as they were shown to be suitable for a FD analysis, especially the MO part \citep{heber15,dumbovic20,belov21}. {The SOHO/EPHIN data were time-shifted to the Earth's bow shock using the OMNIWeb formula for time-shift:}

\begin{equation}
{\Delta t = \frac{X}{V}\cdot\frac{1 + \frac{Y\cdot W}{X}}{1 - \frac{V_e\cdot W}{V}}, }
\label{eq_methods-3.5}
\end{equation}

{\noindent where $X$ and $Y$ are GSE X and Y components of the spacecraft position vector (we used daily positions of SOHO), $V$ is the observed solar wind speed (assumed to be radial), $V_e$ is the speed of the Earth's orbital motion (30 km/s), and $W$=tan(0.5atan ($V$/428)) is the geometry factor (for details, see the OMNIWeb webpage\footnote{\url{https://omniweb.gsfc.nasa.gov/html/ow_data.html}}.}

In order to achieve better visibility of smaller FD variations, SEP events were removed from the data using the $<600$ counts threshold on the B-detector, which  is  sensitive  to  lower  energy  particles, that is, sharp increases due to SEPs. We visually inspected each event and discarded events where either no MO or corresponding FD was observed. In addition, we classified events according to the quality index (QI), subjectively determined by the observer, in the range 1--4, with 1 being the worst quality and 4 being the best quality (see Figure \ref{fig_qi}). The QI is based on criteria regarding the clarity of MO signatures and borders, as well as the clarity of FD signatures and borders. Both are less clear for events designated as having a lower QI. {We note that in QI determination, some more weight was given to the clarity of an onset corresponding to both MO and FD, than to one corresponding to other MO/FD properties, due to its importance for the best-fit application. We also note that although we analysed the clarity of MO borders visually, we did not actually determine or measure MO borders, but FD borders.}

It can be seen in Figure \ref{fig_qi} that the MO signatures are weakest for the QI1 event (bottom right). The magnetic field is increased and the $B_z$ is rotating, although the magnetic field is not very smooth. Plasma beta is low, although the temperature is only partially lower than expected and the expansing profile of the flow speed is missing. The onset of the FD is clear; however, the recovery of the FD seems to be missing. This is likely related to another structure that can be observed trailing the MO. In the QI2 event (bottom left) the MO signatures are clearer. The magnetic field is again increased and the $B_y$ is rotating smoothly. Plasma beta is low, but the temperature is again only partially lower than expected. The flow speed seems to show a globally expanding profile, but with local irregularities. The corresponding FD has a clear onset; however, the onset is slightly shifted in time with respect to the MO onset (according to the low-beta condition). Moreover, the FD shows a double depression without the full recovery to the pre-decrease level. We note that the substructuring of the FD corresponds to the substructuring of the MO.  {The first depression seems to correspond to the rotation of the Bz component from positive to negative values and a small density increase, whereas the second depression seems to correspond to the rotation of the By component from negative to positive values and the low density region. This substructuring might indicate a complex inner structure of the MO, or the passage of the spacecraft under some oblique angle. However, although they both show substructuring, it appears the global structure may correspond to that of a single MO, as was found for example by \citet{nieves-chinchilla18}.} The QI3 event shows clear MO signatures as, in addition to other properties that the QI2 event shows, the flow speed in the QI3 event shows a clearly expanding profile. {Although a depression can be seen to start with the shock-sheath, the second depression of the FD has a clear onset that corresponds to the MO with a clear tendency to return to the pre-decrease level, although it does not (most likely due to the high-speed stream that follows).} Moreover, the FD seems to have different recovery rate at the back (which again corresponds to the substructuring of the MO, similar to the QI2 event, but much less pronounced). Finally, the QI4 event shows clear MO signatures and a clear, symmetric FD with borders that roughly match the FR borders. However, we do note that although globally this event shows clear MO signatures, again some substructuring can be observed. To summarise, in general the best-quality events have nicely observed FD and MO signatures and clear borders. The worst-quality events have unclear borders, pronounced substructures, and/or an unclear recovery. Out of the 30 events in the sample, 6, 12, 8, and 4 have QIs 1, 2, 3, and 4, respectively. The full sample of events is given in Table \ref{tab_sample} in the Appendix \ref{app}.

\subsection{Extraction of FD profiles: The border selection problem}
\label{FD profiles}

In order to apply the best-fit procedure, the FD profiles needed to be extracted, that is, the borders of the FD needed to be selected. {The onset of FD is usually defined as the point where the GCR count starts to drop, whereas the end of FD, in a textbook example, would be the point where it returns to its pre-decrease level \citep[e.g.][]{dumbovic11}. In textbook examples, the recovery of two-step FD is prolonged after the passage of the ICME due to the “shading effect” of the shock-sheath region, whereas for a FR, FD (which can be self-standing or a second decrease in a two-step FD), the end of FD corresponds to the end of the FR \citep{dumbovic20}. However, in many non-textbook cases, the determination of the onset and end of FD is not a trivial procedure} that can be defined by a simple algorithm applicable evenly to all events, as can clearly be seen from Figure \ref{fig_qi}. Therefore, the extraction of FD profiles (from the onset to the end) was performed manually by an observer. This may result in a certain subjectivity when determining the borders, especially the end of FD. Therefore, we applied two different versions of border selection, as is shown in Figure \ref{fig_borders}:
\begin{itemize}
\item \textbf{extended ICME borders} - these encompass typical ICME signatures such as, for example, low temperature or low plasma beta, but with borders set based on the GCR depression onset or end and/or based on the total magnetic field increase onset or end. They thus overlap with typical MO signatures only in a subset of events and they often contain external disturbed parts of the FR.
\item \textbf{inner structure borders} - we find that many events show clearer MO signatures in a sub-interval of what would normally be selected as the global MO structure according to, for example, the low-beta condition. Therefore, first we selected borders to encompass this inner substructure of the CME, which shows clear MO signatures. However, we did this only for events where corresponding clear FD substructure borders could also be found.
\end{itemize}

We find that the inner structure can be identified in 24 events. The inner structure is usually small compared to the typical ICME duration, which is roughly 24 hours \citep{richardson10}. {Consequently, these events have fewer data points to be fitted, but also a more compact profile, without substructures. We did not further refine the QI based on the different border selection, but instead treated the border selection and QI as independent classifications.} In some events the inner structure could not be identified, whereas in some other events the whole duration of the event could be considered as the inner structure. We found four such events, which we in turn regard as not having extended ICME borders. The best-fit procedure is applied to the GCR measurements within the selected borders, and thus slightly different results of best-fit FD amplitude, FD\_bf, can be expected for different border selections. In addition, for each event we determined the FD amplitude using a traditional algorithm-based observational method, FD\_obs, whereby FD was measured from the onset to the minimum. Therefore, this observed FD, FD\_obs, may also be slightly different for different border selections. The borders corresponding to the inner structure and extended ICME are given in Table \ref{tab_sample} in the Appendix \ref{app} for each event, along with the corresponding observed and best-fitted FD.


\section{Results}
\label{results}
\subsection{ForbMod best-fit function testing through synthetic measurements}
\label{testing-synthetic}

\begin{figure*}
\centerline{\includegraphics[width=0.99\textwidth]{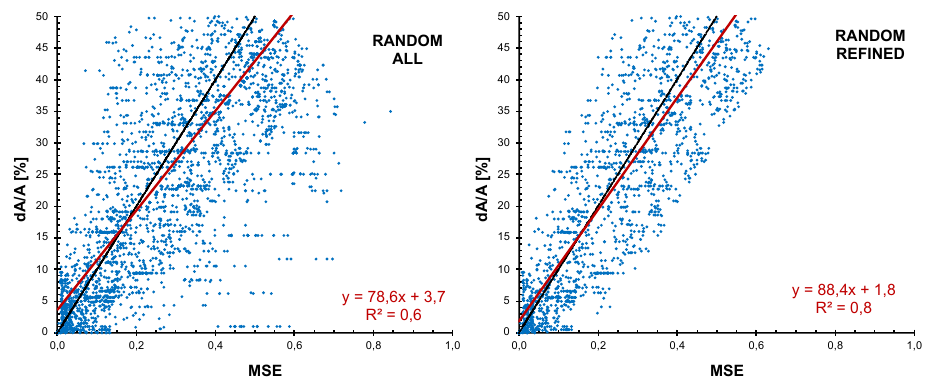}}
\caption{Relative difference of the best-fit FD amplitude to the FD amplitude of the theoretical FD vs $MSE$ for a mixture of different levels of data imperfection effects. In the left plot, no data refinement is included, whereas in the right plot data with extreme deviations from the identity line have been removed (for details, see the main text). The identity line ($y=100*x$) is marked as the solid black line in both figures, whereas the solid red line shows the linear regression line (linear regression equation and squared Pearson's coefficient are given in red in both plots).}
\label{fig_scatter}
\end{figure*}

We applied the ForbMod best-fit procedure to the synthetic data in order to check how the goodness of fit ($MSE$) behaves for each of the different effects, but also for different levels of a certain effect. For each best-fit to the synthetic data we calculated $MSE$ and the relative difference of the best-fit curve minimum to the minimum of the theoretical curve (i.e. the relative difference of the best-fit FD amplitude to the FD amplitude of the theoretical FD, $\mathrm{d}A/A$). We first applied the ForbMod best-fit procedure to synthetic data with different levels of Y-noise and X-noise. The results are shown in Figure \ref{fig_mse} a and b. We can see that, for the same level of Y-noise, $\mathrm{d}A/A$ and $MSE$ are correlated, whereas for the same level of X-noise the $\mathrm{d}A/A$ remains constant. Therefore, by increasing the X-noise we increase both $MSE$ and $\mathrm{d}A/A$. However, by increasing the Y-noise we only increase $MSE$, and $\mathrm{d}A/A$ does not change drastically. This is related to the fact that Y-noise is applied in the same direction in which the $MSE$ is calculated. Simply put, the random noise up and down around the curve will contribute to the $MSE$, but the average curve will stay the same as the original one.

We next applied the ForbMod best-fit procedure to synthetic data with different levels of X-noise and asymmetry. This is shown in Figure \ref{fig_mse}c. We can see that for lower values of asymmetry ($<0.15\%$) asymmetry works in the same direction as the X-noise: the larger the $MSE$, the larger the $\mathrm{d}A/A$. However, for large asymmetry there seems to be a saturation of $\mathrm{d}A/A$, which remains scattered around 60\% regardless of the $MSE$. We note that combining three different data imperfection effects (Y-noise, X-noise, and asymmetry) very quickly leads to large $\mathrm{d}A/A$, as the data does not resemble the FD anymore. Therefore, it is already impossible to apply the best-fit procedure to data with Y-noise, X-noise, and asymmetry greater than 15\%, 15\%, and 0.15\%, respectively.

We next applied the data gap to the synthetic measurements and repeated the best-fit procedure. We applied a data gap 10\% in size to five different positions, {starting from the onset to the minimum, as is described in Section \ref{synthetic}. The results are shown in Figure \ref{fig_mse}d), where we find} that for a data gap position closer to the FD minimum, the increase in $MSE$ results in an increase in $\mathrm{d}A/A$ greater than for data gaps closer to the onset of FD. In other words, we might expect to fit worse when our dataset has a gap close to the FD minimum. We also checked how the data size influences the best-fit procedure (Figure \ref{fig_mse}e). We find that for gap sizes of up to roughly 35\% the scatter of the $MSE$ for the same $\mathrm{d}A/A$ increases - depending on where the data gap is and what other noise the data has, the data gap of the same size can lead to a range of different $\mathrm{d}A/A$, and this range increases with the gap size. However, for gap sizes $>35\%$ we observe the opposite trend. This is related to the fact that with an increase in the data gap we removed data and with it the other data imperfection effects.

Finally, we applied different combinations of data imperfection effects using three different resolutions of the original theoretical curve data (2000, 200, and 20 data points). In Figure \ref{fig_mse}f we can see that for smaller resolutions we expect a larger scatter of $MSE$ for the same $\mathrm{d}A/A$.

\begin{figure*}
\centerline{\includegraphics[width=0.99\textwidth]{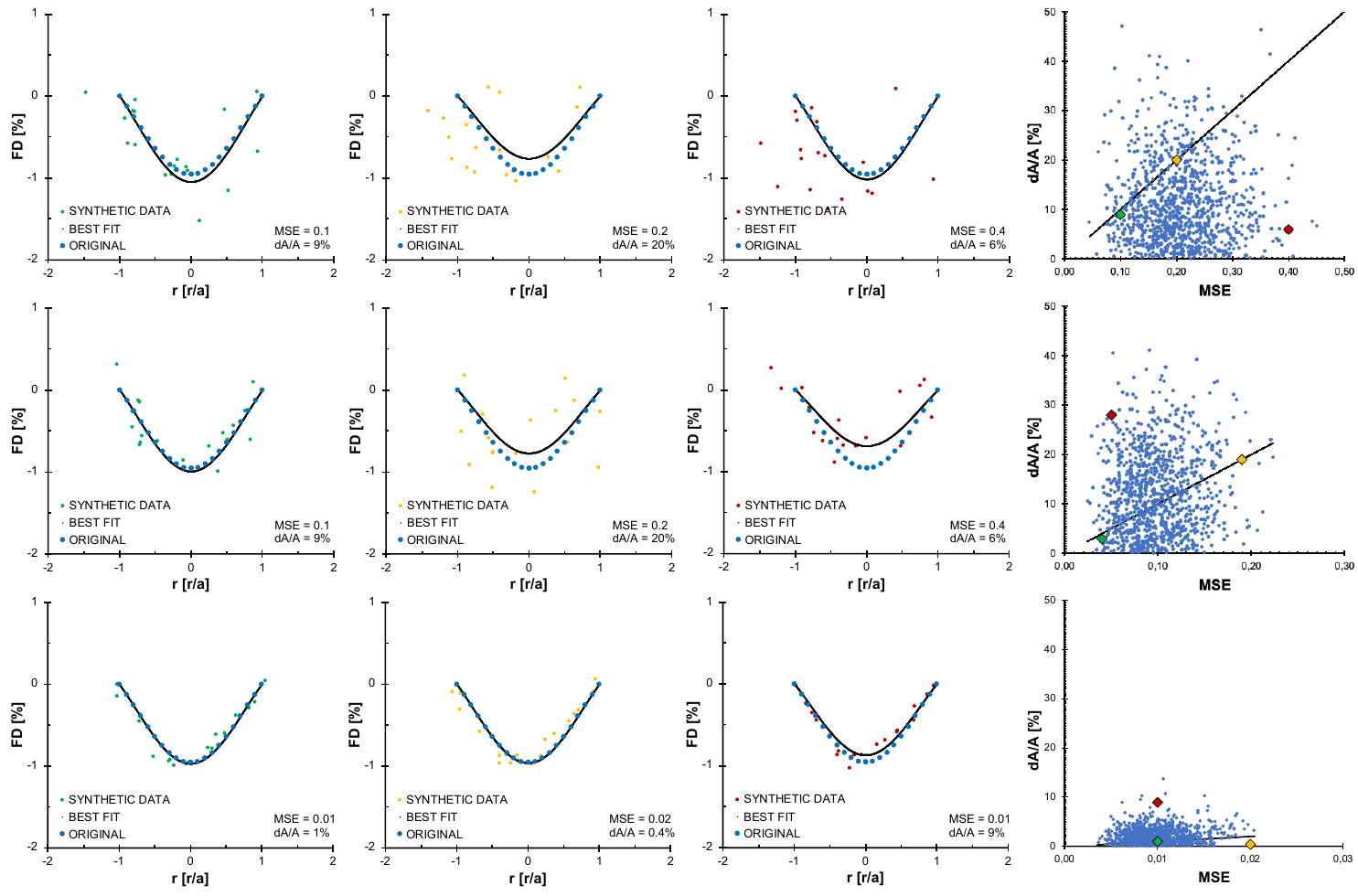}}
\caption{Synthetic data and the best-fit results for several selected examples. Different examples correspond to different combinations of data imperfection effects, where the set of values of the data imperfection effects is different for different rows and varies from high to low and top to bottom, respectively: 1) top: Y-noise=30\%, X-noise=30\%, asym=0.15\%, gap position=3, gap size=30\%, data resolution=20pts; 2) middle: Y-noise=20\%, X-noise=20\%, asym=0.10\%, gap position=2, gap size=20\%, data resolution=20pts; 3) bottom: Y-noise=5\%, X-noise=5\%, asym=0.05\%, gap position=2, gap size=10\%, data resolution=20pts. The right-most plot in each row shows the scatter of different $\mathrm{d}A/A$ and $MSE$ values obtained through 1000 runs for the same maximum data imperfection effect levels. Different combinations of data imperfection effects may lead to different $MSE$ and $\mathrm{d}A/A$ behaviour. The three examples shown left of the corresponding scatter plot are highlighted in a colour corresponding to the colour of the synthetic data for the given event (green for the left-most, yellow for the middle, and red for the right-most examples). Blue data points correspond to the original theoretical curve, and the best-fit function is given by the black curve.}
\label{fig_examples}
\end{figure*}

In Figure \ref{fig_scatter} we show the relative difference of the best-fit FD amplitude to the FD amplitude of the theoretical FD versus the $MSE$ for a mixture of different levels of data imperfection effects. Combinations of different imperfection effects can quickly lead to very high $MSE$ and $\mathrm{d}A/A$, and therefore we constrained combinations to the following levels: Y-noise=15\%, X-noise=15\%, asym=0.15\%, and gap size=30\%. However, we allowed parameters to go beyond these levels when one or more of the data imperfection parameters was kept at level 0. We can see in Figure \ref{fig_scatter} that the linear fit is not far away from the identity line ($y=100*x$); however, the scatter of the data is quite large.

The linear fit moves even closer to the identity line when we refine the sample by removing extremely outlying points, that is, those that have $\mathrm{d}A/A>20\%$ away from the identity line and extremely low or high corresponding $MSE$, that is, $<0.2$/$>0.3$, respectively. This is shown in the right plot of Figure \ref{fig_scatter}. Therefore, we next analysed visually the synthetic data and the best-fit function, as well as the $MSE$ and $\mathrm{d}A/A$ for cases which are that to the identity line and those far away. This is shown for nine selected examples in Figure \ref{fig_examples}. The examples are given in the nine panels on the left of the Figure, whereas the right-most plots show the scatter plots of $\mathrm{d}A/A$ vs. $MSE$ for three different sets of randomly produced synthetic data. The scatter plots differ slightly due to randomness of producing synthetic data, but in general show similar behaviour. The three examples shown left of the corresponding scatter plot are highlighted in a colour corresponding to the colour of the synthetic data for the given event (green for the left-most, yellow for the middle, and red for the right-most examples). For synthetic data of the same colour, top-to-bottom panels show different levels of data imperfection effects, ranging from highest (upper plots) to lowest (bottom plots).

We can immediately notice that the level of data imperfection affects the visual representation of the FD. The top plots showing very high levels of data imperfection effects have a barely identifiable FD, whereas in the bottom plots the data depict an almost ideal symmetric FD profile. This visual representation is in principal reflected by the $MSE$ -- lower plots have much smaller $MSE$ compared to the upper plots. The middle plots present intermediate cases between extreme examples. Looking at the difference between the theoretical and best-fit curve, $\mathrm{d}A/A$, we can see that it is negligible for the bottom left-most plot, whereas in the upper left-most plot it almost reaches 10\%. However, for the top-to-bottom right-most examples it is reversed -- $\mathrm{d}A/A$ is small for the top right example, whereas in the bottom right-most example it almost reaches 10\%. This reflects the previously observed fact that for some data imperfection effects $MSE$ is not correlated with $\mathrm{d}A/A$. These are the cases that are far away from the identity line. We note that in the example of low levels of the data imperfection effects (bottom plots) $MSE$ is low ($<0.03$) in all three cases and $\mathrm{d}A/A$ is within 10\%. Therefore, we do not regard any of the bottom plots cases as being far away from the identity line.

From this analysis we see that, if we exclude events for which visual representation is not clear enough to identify FD, we are left with events close to the identity line. If we further restrict ourselves only to events with $MSE<0.1$, we are left with events with $\mathrm{d}A/A<$10\%. Therefore we can conclude that $MSE<0.1$ can serve as a goodness-of-fit measure resulting in  $<10\%$ error of the fitted amplitude under the following assumptions: 1) the real data behave as synthetic data; 2) the data in general depict the behaviour of the theoretical ForbMod FD (with some levels of data noise); 3) FD is clearly visible in the data and not masked by noise.

\subsection{ForbMod best-fit function application to SOHO/EPHIN measurements}
\label{testing-ephin}

\begin{figure*}
\centerline{\includegraphics[width=0.9\textwidth]{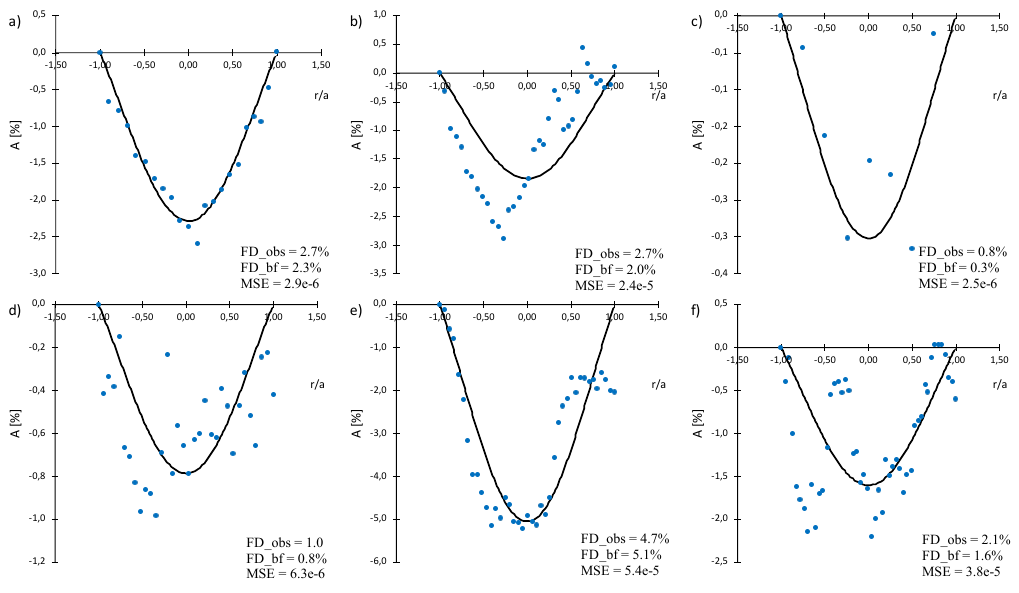}}
\caption{Six types of ForbMod fits to EPHIN data based on visual analysis: a) a good fit (event1, inner structure borders); b) an asymmetric fit (event1, extended ICME borders); c) a few-data-points fit (event2, inner structure borders); d) a large-scatter fit (event2, extended ICME borders); e) an under-recovery fit (event4, extended ICME borders); and f) a substructuring fit (event17, extended ICME borders). FD\_obs was FD amplitude-determined using a traditional algorithm-based observational method, whereby FD was measured from the onset to the minimum. FD\_bf is the FD amplitude obtained from the ForbMod best-fit procedure and MSE is the mean standard error.}
\label{fig_profiles}
\end{figure*}

We next performed the ForbMod best fit on the sample of the events, as is described in Section \ref{sample}. We first analysed the fits visually and found we could divide the fits into six types:
\begin{itemize}
\item \textbf{a good fit} - where there was visually a good match between data points and the best-fit function
\item \textbf{an asymmetric fit} - where the fitted function was shifted along the x axis compared to the data points
\item \textbf{a few-data-points fit} - where the fit was performed on only a small number of data points
\item \textbf{a large-scatter fit} - where the fit was performed on data points with a large scatter
\item \textbf{an under-recovery fit} - where the data points did not return to the pre-decrease level
\item \textbf{a substructuring fit} - where a single fit was performed while the data points indicated possible substructuring
\end{itemize}

Examples of fit types are given in Figures \ref{fig_profiles}a-f. However, we note that some events may fall into more than one category; for example, an event might show under-recovery as well as substructuring. As can be seen in Table \ref{tab_sample} all but one good fit events have inner structure borders. They also on average have the smallest MSE (3.9e-5). A few-data-points fit is another type present only for inner structure borders events. On the other hand, the large-scatter fit and the substructuring fit seem to be categories present predominantly for extended ICME borders and on average have the largest MSEs (1.8e-4 and 1.5e-4, respectively). We note that for all of the events $MSE<0.01$. Therefore, although some profiles do not show a visually pleasing FD, the ForbMod best-fit function still manages to find a solution with a relatively small mean standard error (MSE). 

We next analysed best-fit results for four different QIs, as is described in Section \ref{sample} (see Table \ref{tab_qi}). We do not find notable differences between events marked by different QIs, in either the difference between FD\_bf and FD\_obs, the values of MSE, or the spread of MSE. Surprisingly, we find that MSE on average takes the smallest values for QI 1, i.e. the least clear events. This is likely related to the fact that these events also on average have the smallest FD amplitudes, both FD\_bf and FD\_obs (1.1\% and 1.2\%, respectively). 

\begin{table}
\caption{Statistics for the best-fit amplitude, FD\_bf, the amplitude obtained using a traditional algorithm-based observational method, FD\_obs, and the MSE for four different QIs (as is described in Section \ref{sample}).}
\label{tab_qi}
\centering
\small
\begin{tabular}{lcccc} 
\hline
                        &       QI 1            &       QI 2            &       QI 3               &       QI 4            \\
\hline
No. of events   &       6               &       12              &       8               &       4               \\
\hline
\multicolumn{5}{l}{inner structure}                                                                     \\
No. of events   &       4               &       10              &       7               &       3               \\
FD\_obs mean    &       1,2             &       3,0             &       4,0             &       3,5             \\
FD\_bf mean     &       1,1             &       2,8             &       3,6             &       3,1             \\
MSE mean        &       7,8E-06 &       5,0E-05 &       5,7E-05 &       2,7E-05 \\
MSE stdev       &       4,7E-06 &       6,0E-05 &       4,2E-05 &       3,3E-05 \\
\hline
\multicolumn{5}{l}{extended ICME structure}                                                     \\
No. of events   &       6               &       10              &       6               &       3               \\
FD\_obs mean    &       1,8             &       3,6             &       4,0             &       5,3             \\
FD\_bf mean     &       1,7             &       2,2             &       2,6             &       5,2             \\
MSE mean        &       4,8E-05 &       1,1E-04 &       1,3E-04 &       2,0E-04 \\
MSE stdev       &       4,7E-05 &       1,1E-04 &       9,8E-05 &       2,7E-04 \\
\hline
\end{tabular}
\end{table}

\begin{figure*}
\centerline{\includegraphics[width=0.8\textwidth]{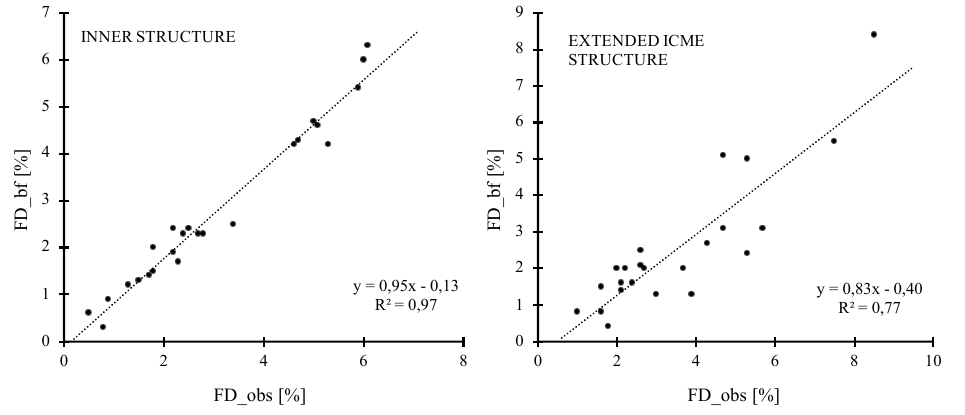}}
\caption{Linear regression between the best-fitted FD amplitude, FD\_bf, and the FD amplitude measured using the traditional algorithm-based observational method (from the onset to the minimum), FD\_obs, for two different border selections (inner structure borders and extended ICME borders)}
\label{fig_corr}
\end{figure*}

We next compared the calculated best-fit amplitude, FD\_bf, with the amplitude obtained using a traditional algorithm-based observational method, FD\_obs, whereby FD\_obs was measured from the onset to the minimum. For the inner structure borders, we find the average FD\_bf of 2.8\%, which is somewhat smaller than the average FD\_obs (3.1\%). We note that this difference is expected, because the traditional algorithm-based observational method measures maximum amplitude, without taking into account possible scatter of the data. For extended ICME borders we find an even larger difference between the two, with an average FD\_bf of 2.5\% and an average FD\_obs of 3.5\%. This reflects the fact that the scatter of the data that was fitted is on average larger for extended ICME borders compared to inner structure borders, as can be seen from the average values of MSE being 1.1e-4 and 4.2e-5, respectively.

We also performed a linear regression and calculated the Pearsons correlation coefficient for FD\_bf  versus FD\_obs separately for inner structure borders and extended ICME borders. The results are presented in Figure \ref{fig_corr} and confirm the findings above. We observe a negative intercept in both linear regressions, confirming a systematically smaller FD\_bf compared to FD\_obs. The larger negative intercept for extended ICME borders indicates a larger difference between FD\_bf and FD\_obs. In addition, we can see that the Pearsons correlation coefficient (i.e. its squared value given by $R^2$) is smaller for extended ICME borders related to the larger scatter. However, we do note that the correlation is high for both border selection cases  ($R>0.8$). Thus, we can conclude that the ForbMod best-fit procedure performs similarly to the traditional algorithm-based observational method, but with slightly smaller values for the FD amplitude, as it's taking into account the noise in the data. We note that in two events FD\_obs could not be determined because of the data gap at the onset of the FD (see Table \ref{tab_sample}). In those cases, FD\_bf could still be determined, which is an obvious advantage of the best-fit procedure. 

\section{Summary and conclusions}
\label{conclusion}

We developed a new method to measure the FD amplitude, which is based on the analytical diffusion-expansion FD model for FRs (ForbMod). The ForbMod best-fit procedure we developed performs a non-linear fitting of the Bessel function of the zero order (i.e. the radial part of the ForbMod solution) to the designated time series. The non-linear fitting was performed by calculating all possible ForbMod curves constrained within FR borders to the designated dataset and minimising the MSE. The input for the procedure is the designated dataset with defined FR borders (i.e. the dataset translated into the FR coordinate system radial direction, with the centre of the FR at 0 and FR borders at +/-1, respectively). The output of the procedure is the best-fitted amplitude of the measured FD and MSE.

In order to evaluate the performance of the ForbMod best-fit procedure, we used synthetic measurements. Synthetic measurements were produced by calculating the theoretical ForbMod curve for a specific example CME and then applying various effects to the data to mimic the imperfection of the real measurements. Using synthetic measurements we could test how far away the best-fitted function is from the real one for different cases. We found that if the FD is clearly visible in the data the reliable ForbMod best-fit function should result in the MSE being <0.1 with an expected <10\% relative error in the fitted amplitude, roughly along the identity line.

We next tested the ForbMod best-fit function on the real data, measured by detector F of the SOHO/EPHIN instrument. We used a list of SOHO/EPHIN detected FDs from the time period 1995--2015 and a catalogue of near-Earth ICMEs to compile a list of events suitable for analysis. The sample contains 30 events, all of which have a distinct FD corresponding to the magnetic obstacle (MO). We categorised the events according to different QIs, which mark different levels of clarity of the events. We extracted FD profiles for each event in the sample from the onset to the end of FD and prepared them for the application of the ForbMod best-fit function. The extraction of FD profiles (from the onset to the end) was performed manually by an observer, whereby we applied two different versions of border selection: 1) the inner substructure of the CME, which shows clear MO signatures, and 2) the extended ICME structure, which may also encompass outer, disturbed layers of the ICME. 

We do not find notable differences between events marked by different QIs. We compared the calculated best-fit amplitude with the amplitude obtained using a traditional algorithm-based observational method (measured from the onset to the minimum) and found a good match between the two. The ForbMod best-fit procedure performs similarly to the traditional algorithm-based observational method, but with slightly smaller values for the FD amplitude, as it takes into account the noise in the data. For events with two different versions of border selection we find that the best-fit applied on extended ICME structure borders results in slightly larger MSE and differences compared to the traditional method due to the larger scatter of the data points. We find that the best-fit results can visually be categorised into six different FD profile types. Although some profiles do not show a visually pleasing FD, the ForbMod best-fit function still manages to find a solution with a relatively small MSE.

Finally, we find that the best-fit procedure has an advantage compared to the traditional method as it can estimate the FD amplitude even when there is a data gap at the onset of the FD. The method could be further advanced by taking into account best-fitting for multiple-border selection, and thus reducing the subjective impact of the observer. {With further improvements in the direction of border selection, the method would not only advance the estimation of the magnitude of the FD, but also the determination of its onset, end, and consequently duration. Such an application would provide the means for a quick, objective, reliable, and robust way of measuring FDs and may eventually facilitate the automation of FD measurements.}

\begin{acknowledgements}
This research was performed in the scope of the ESA funded project ``Forbush decrease analysis using model fitting to SOHO/EPHIN data (ForbMod)'' (ESA contract No. 4000135501 lz1lNL/SC/hm). M.D. acknowledges support by the Croatian Science Foundation under the project IP-2020-02-9893 (ICOHOSS). We acknowledge the OMNIWeb database as the source of data used, as well as the SOHO/EPHIN. SOHO/EPHIN is supported by the Ministry of Economics via DLR grant 50OG1702. The OMNI data were obtained from the GSFC/SPDF OMNIWeb interface at https://omniweb.gsfc.nasa.gov.
\end{acknowledgements}
%
%
\bibliographystyle{aa} 
\bibliography{REFs.bib}

%
%
\begin{appendix} 
\section{The sample}
\label{app}

\begin{landscape}
\begin{table}
\caption{Full sample of analysed ICMEs and FDs}
\label{tab_sample}
\centering
\small
\begin{tabular}{|ccc|cccccc|cccccc|} 
\hline           
event   &       ICME start      &       QI      &       \multicolumn{6}{c|}{inner structure}                                                                                      &       \multicolumn{6}{c|}{extended ICME}                                                                                   \\
No.             &       date                    &               &       DOY start   &       DOY end &       FD\_obs &       FD\_bf  &       MSE             &       fit type    &       DOY start       &       DOY end &       FD\_obs &       FD\_bf  &       MSE             &       fit type    \\
\hline
1               &       2008-09-17      &       4       &       261,2   &       262,1   &       2,7             &       2,3             &       2,9E-06 &       1               &       261,1   &       262,6   &       2,7             &       2,0             &       2,4E-05 &       2               \\
2               &       2009-09-30      &       1       &       273,3   &       273,6   &       0,8             &       0,3             &       2,5E-06 &       3               &       273,1   &       274,4   &       1,0             &       0,8             &       6,3E-06 &       4               \\
3               &       2009-11-14      &       2       &       318,5   &       318,9   &       0,5             &       0,6             &       1,9E-06 &       3               &       317,9   &       319,2   &       2,6             &       2,5             &       2,1E-05 &       5, 6               \\
4               &       2010-02-07      &       4       &       38,8            &       40,0            &       5,1             &       4,6             &       6,4E-05 &       1               &       38,7            &       40,4            &       4,7             &       5,1             &       5,4E-05 &       5               \\
5               &       2010-04-05      &       4       &       N/A             &       N/A             &       N/A             &       N/A             &       N/A             &       N/A             &       95,4            &       96,8            &       8,5             &       8,4             &       5,1E-04 &       2, 5, 6    \\
6               &       2010-05-28      &       2       &       148,8   &       149,1   &       5,0             &       4,7             &       1,6E-04 &       3               &       148,6   &       149,9   &       7,5             &       5,5             &       3,0E-04 &       5, 6               \\
7               &       2010-06-21      &       2       &       172,3   &       172,6   &       2,2             &       1,9             &       1,1E-05 &       3, 5               &       172,1   &       173,8   &       3,0             &       1,3             &       3,3E-05 &       6               \\
8               &       2010-12-20      &       1       &       354,4   &       355,0   &       0,9             &       0,9             &       5,5E-06 &       3               &       353,7   &       355             &       1,6             &       1,5             &       4,1E-05 &       5, 6               \\
9               &       2011-03-30      &       2       &       88,9            &       90,0            &       5,9             &       5,4             &       1,4E-04 &       5               &       88,9            &       91,7            &       5,7             &       3,1             &       2,8E-04 &       2               \\
10              &       2011-09-17      &       3       &       260,6   &       262,0   &       6,1             &       6,3             &       8,0E-05 &       1               &       N/A             &       N/A             &       N/A             &       N/A             &       N/A             &       N/A             \\
11              &       2012-10-31      &       1       &       N/A             &       N/A             &       N/A             &       N/A             &       N/A             &       N/A             &       305,9   &       307,3   &       N/A             &       3,0             &       1,4E-04 &       2               \\
12              &       2012-11-26      &       2       &       331,6   &       332,2   &       1,7             &       1,4             &       3,0E-06 &       1               &       331,5   &       332,5   &       1,6             &       0,8             &       4,1E-05 &       4               \\
13              &       2013-02-16      &       1       &       48,3            &       48,9            &       1,3             &       1,2             &       1,3E-05 &       2               &       47,8            &       49,8            &       2,1             &       1,4             &       1,8E-05 &       4, 6               \\
14              &       2013-04-30      &       1       &       N/A             &       N/A             &       N/A             &       N/A             &       N/A             &       N/A             &       120,5   &       121,3   &       2,0             &       2,0             &       4,3E-05 &       4, 5               \\
15              &       2013-06-06      &       3       &       157,6   &       158,4   &       4,6             &       4,2             &       4,7E-05 &       5               &       157,6   &       159             &       5,3             &       5,0             &       2,7E-04 &       5, 6               \\
16              &       2013-06-27      &       3       &       179,1   &       179,6   &       5,3             &       4,2             &       1,2E-04 &       3               &       179,1   &       180,3   &       5,3             &       2,4             &       1,9E-04 &       6               \\
17              &       2013-07-05      &       1       &       186,7   &       187,3   &       1,8             &       2,0             &       1,0E-05 &       1               &       186,7   &       188,7   &       2,1             &       1,6             &       3,8E-05 &       6               \\
18              &       2013-07-12      &       3       &       194,1   &       195,2   &       4,7             &       4,3             &       9,5E-05 &       2, 5               &       194,1   &       196,0   &       4,7             &       3,1             &       2,0E-04 &       2, 5, 6    \\
19              &       2013-11-11      &       2       &       315,7   &       316,2   &       1,5             &       1,3             &       1,5E-05 &       3               &       315,5   &       316,5   &       3,9             &       1,3             &       1,2E-04 &       4               \\
20              &       2013-12-15      &       2       &       349,7   &       350,2   &       6,0             &       6,0             &       1,0E-04 &       1               &       N/A             &       N/A             &       N/A             &       N/A             &       N/A             &       N/A             \\
21              &       2014-04-29      &       3       &       119,9   &       120,7   &       2,2             &       2,4             &       1,6E-05 &       1               &       119,8   &       120,9   &       2,6             &       2,1             &       3,7E-05 &       2               \\
22              &       2014-05-30      &       4       &       150,6   &       152,3   &       2,8             &       2,3             &       1,3E-05 &       5               &       N/A             &       N/A             &       N/A             &       N/A             &       N/A             &       N/A             \\
23              &       2014-08-19      &       3       &       231,7   &       232,2   &       1,8             &       1,5             &       2,6E-05 &       3               &       231,7   &       233,8   &       1,8             &       0,4             &       3,9E-05 &       4               \\
24              &       2016-04-14      &       3       &       105,3   &       106,0   &       3,4             &       2,5             &       1,4E-05 &       5               &       N/A             &       N/A             &       N/A             &       N/A             &       N/A             &       6               \\
25              &       2016-07-24      &       2       &       N/A             &       N/A             &       N/A             &       N/A             &       N/A             &       N/A             &       206,7   &       207,7   &       N/A             &       1,6             &       6,8E-06 &       1               \\
26              &       2016-10-12      &       3       &       N/A             &       N/A             &       N/A             &       N/A             &       N/A             &       N/A             &       287             &       288,7   &       4,3             &       2,7             &       7,2E-05 &       4               \\
27              &       2016-11-09      &       2       &       N/A             &       N/A             &       N/A             &       N/A             &       N/A             &       N/A             &       314,7   &       315,7   &       2,2             &       2,0             &       4,2E-05 &       6               \\
28              &       2017-04-13      &       2       &       104             &       104,6   &       2,3             &       1,7             &       2,1E-05 &       2               &       N/A             &       N/A             &       N/A             &       N/A             &       N/A             &       N/A             \\
29              &       2018-08-25      &       2       &       237,5   &       238,2   &       2,4             &       2,3             &       3,0E-05 &       6               &       237,5   &       238,4   &       2,4             &       1,6             &       1,0E-04 &       2               \\
30              &       2019-05-13      &       2       &       134,2   &       134,7   &       2,5             &       2,4             &       1,9E-05 &       5               &       134             &       136,0   &       3,7             &       2,0             &       1,5E-04 &       6               \\
\hline
\multicolumn{15}{|l|}{ } \\
\multicolumn{15}{|l|}{The ICME start date corresponds to the first detected ICME signature, which may be shock, sheath, pile-up, or MO (depending on the event).} \\
\multicolumn{15}{|l|}{The QI marks the clarity of MO and FD signatures and borders and is subjectively determined by the observer, as is explained in Section \ref{sample}.} \\
\multicolumn{15}{|l|}{Measurements are separated for the inner structure and extended ICME, as is explained in Section \ref{FD profiles}.} \\
\multicolumn{15}{|l|}{For each, we determined the onset and end of the FD using the day of the year (DOY), DOY start, and DOY end, respectively;} \\
\multicolumn{15}{|l|}{FD amplitude measured using the traditional method, FD\_obs; best-fitted FD amplitude, FD\_bf; MSE;} \\
\multicolumn{15}{|l|}{and fit type as visually determined by the observer (for details see Section \ref{testing-ephin}).} \\
\multicolumn{15}{|l|}{ } \\
\hline
\end{tabular}
\end{table}
\end{landscape}

\end{appendix}
\end{document}